\documentclass[aps,pre,amsmath,lengthcheck,superscriptaddress]{revtex4-2}

\usepackage{graphicx}\graphicspath{ {figures/} }
\usepackage{hyperref}
\hypersetup{colorlinks,allcolors=blue,breaklinks}

\newcommand{\be}{\begin{equation}}
\newcommand{\ee}{\end{equation}}
\newcommand{\ba}{\begin{align}}
\newcommand{\ea}{\end{align}}
\newcommand{\bi}{\begin{itemize}}
\newcommand{\ei}{\end{itemize}}

\newcommand{\tr}[1]{\mathrm{tr}\left\{#1\right\}}

\newcommand{\la}{\left\langle}
\newcommand{\ra}{\right\rangle}
\newcommand{\pd}{\partial}

\newcommand{\bla}{bla\\bla\\bla\\bla\\bla}

\newcommand{\mc}[1]{\mathcal{#1}}

\begin{document}

\title{Analytical shortcuts to adiabaticity of weakly driven processes}

\author{Pierre Naz\'e}
\email{pierre.naze@icen.ufpa.br}

\affiliation{\it Universidade Federal do Par\'a, Faculdade de F\'isica, ICEN,
Av.~Augusto Corr\^ea, 1, Guam\'a, 66075-110, Bel\'em, Par\'a, Brazil}

\date{\today}

\begin{abstract}

The analytical expression for shortcuts to adiabaticity for any switching time and any thermally isolated system performing a finite-time and weakly driven process is presented. It is based on the analytical solution of the optimal protocols of weak processes for open systems. The extension to adiabatic processes was made by employing the concept of waiting time. The shortcuts to adiabaticity are proven by showing that the excess power is null in all instants, indicating that no nonequilibrium excitation occurs along the driving. Two examples are solved to verify the validity of such shortcuts: the typical case of oscillatory relaxation function and the transverse-field quantum Ising chain. In this last case, it is shown that non-quenching process outperforms the quenching one by suppressing the non-equilibrium excitations until the critical point is not achieved. 

\end{abstract}

\maketitle

\section{Introduction}
\label{sec:introduction}

Although finding finite-time optimal protocols for the work spent in a thermodynamic process may be urgent, there is also interest in identifying situations where these protocols are the best ones. For thermally isolated systems, this kind of protocols is called {\it shortcut to adiabaticity}, where a finite-time driving makes the adiabatic invariant constant, and the work spent is the quasistatic work~\cite{chen2010fast, guery2019shortcuts, torrontegui2013shortcuts, delcampo2013shortcuts, deffner2014classical, an2016shortcuts, takahashi2013transitionless, campbell2015shortcut, saberi2014adiabatic, patra2017shortcuts}. Recently it has been demonstrated that such quasistatic work outperforms any other~\cite{naze2025fluctuation}.

Previous studies show that such shortcuts to adiabaticity exist in the context of finite-time and weak drivings~\cite{acconcia2015degenerate,acconcia2015shortcuts}. However, the proposed solutions were restricted to continuous protocols, therefore removing a class of admissible functions which admits distributional solutions involving Dirac deltas and their derivatives~\cite{kirk2004optimal}. In another study, universal solutions for the optimal protocol for isothermal and weak processes were found considering these admissible functions~\cite{naze2024analytical}. The question is then proposed: can one extend such a procedure to adiabatic processes and find universal shortcuts to adiabaticity where these admissible functions are considered?

The answer is positive. First, after proposing the concept of waiting time to unify the treatment of the optimal protocols for isothermal and adiabatic processes, I find a structure for such a shortcut to adiabaticity, valid for any switching time and any thermally isolated system. Also, the proof that the excess power is null for all instants is made, showing that no non-equilibrium excitations occur during the driving. Two examples are solved that illustrate the result.

In particular, one of these examples is the transverse-field quantum Ising chain, which is a prototype of an adiabatic quantum computer~\cite{deffner2019quantum,Morita2008,Hauke2020,Chakrabarti2023,Hegde2022,Khezri2022,King2022,Soriani2022,Yulianti2022,soriani2022assessing,farhi2000quantum, farhi2001quantum, albash2018adiabatic, das2008colloquium, aharonov2007adiabatic, roland2002quantum, childs2001robustness, lidar2008towards, amin2009decoherence, jansen2007bounds, hen2015probing}. Treating the particular quenching process case, where the waiting time diverges, I show that the non-quenching case outperforms the first by suppressing any nonequilibrium excitations of the system along the process. The optimal protocol of the quenching case turns out to be the pausing protocol, already reported in the literature~\cite{marshall2019power, albash2020comparing, chen2022advantage, zielewski2022method, chen2023efficiency, chen2020why, calude2017adiabatic, marsh2019enhancing, passarelli2020reverse}.

\textcolor{black}{Despite significant advances in the development of shortcuts to adiabaticity, many existing approaches remain limited in scope. Most notably, shortcuts to adiabaticity protocols derived from counter-diabatic driving or invariant-based inverse engineering often rely on smooth and continuous control functions, excluding protocols that may contain distributional components such as Dirac deltas and their derivatives. These constraints restrict the range of implementable solutions, particularly in experimental scenarios where singular driving schemes may naturally arise.}

\textcolor{black}{The analytical framework developed here provides a unifying and general solution for shortcuts to adiabaticity in thermally isolated systems subjected to weak driving, valid for any switching time. By introducing the concept of waiting time, this work bridges the gap between isothermal and adiabatic processes and enables a unified description of optimal protocols that minimize irreversible excitations. Unlike existing techniques that are limited to smooth drivings, the present solution includes optimal protocols containing singular terms --- thus extending the class of admissible control functions. This inclusion is not merely formal; it enables exact suppression of non-equilibrium excitations even near critical points, which is particularly relevant for applications in quantum annealing and adiabatic quantum computing~\cite{deffner2019quantum,Morita2008,Hauke2020,Chakrabarti2023,Hegde2022,Khezri2022,King2022,Soriani2022,Yulianti2022,soriani2022assessing,farhi2000quantum, farhi2001quantum, albash2018adiabatic, das2008colloquium, aharonov2007adiabatic, roland2002quantum, childs2001robustness, lidar2008towards, amin2009decoherence, jansen2007bounds, hen2015probing}.}

\section{Preliminaries}
\label{sec:preliminaries}

I start by defining notations and developing the main concepts to be used in this work. \textcolor{black}{This technical introductory section is based in Ref.~\cite{naze2023adiabatic,naze2024analytical}}.

Consider a quantum system with a Hamiltonian $\mc{H}(\lambda(t))$, where $\lambda(t)$ is a time-dependent external parameter. Initially, this system is in contact with a heat bath of temperature $\beta\equiv {(k_B T)}^{-1}$, where $k_B$ is Boltzmann's constant. The system is then decoupled from the heat bath and, during a switching time $\tau$, the external parameter is changed from $\lambda_0$ to $\lambda_0+\delta\lambda$. The average work performed on the system during this process is
\be
W \equiv \int_0^\tau \la\pd_{\lambda}\mc{H}(t)\ra\dot{\lambda}(t)dt,
\label{eq:work}
\ee
where $\partial_\lambda$ is the partial derivative for $\lambda$ and the superscripted dot is the total time derivative. The generalized force $\la\pd_{\lambda}\mc{H}(t)\ra$ is calculated using the trace over the density matrix $\rho(t)$
\be
\la A(t)\ra =\tr{A\rho(t)}
\ee
where $A$ is some observable. The density matrix $\rho(t)$ evolves according to Liouville equation
\be
\dot{\rho} = -\frac{1}{i\hbar}[\rho,\mc{H}],
\ee
where $[\cdot,\cdot]$ is the commutator and $\rho(0)=\rho_c$ is the initial canonical density matrix. Consider also that the external parameter can be expressed as
\be
\lambda(t) = \lambda_0+g(t)\delta\lambda,
\ee
where to satisfy the initial conditions of the external parameter, the protocol $g(t)$ must satisfy the following boundary conditions
\be
g(0)=0,\quad g(\tau)=1. 
\label{eq:bc}
\ee

Linear response theory aims to express the average of some observable until the first order of some perturbation considering how this perturbation affects the observable and the non-equilibrium density matrix \cite{kubo2012}. In our case, we consider that the parameter does not considerably changes during the process, $|g(t)\delta\lambda/\lambda_0|\ll 1$, for all $t\in[0,\tau]$. Using the framework of linear-response theory \cite{kubo2012}, the generalized force $\la\pd_{\lambda}\mc{H}(t)\ra$ can be approximated until the first-order as
\begin{equation}
\begin{split}
\la\pd_{\lambda}\mc{H}(t)\ra =&\, \la\pd_{\lambda}\mc{H}\ra_0-\delta\lambda\widetilde{\Psi}_0 g(t)\\
&+\delta\lambda\int_0^t \Psi_0(t-t')\dot{g}(t')dt',
\label{eq:genforce-relax}
\end{split}
\end{equation}
where 
\be
\Psi_0(t)=\beta\langle\partial_\lambda\mathcal{H}(t)\partial_\lambda\mathcal{H}(0)\rangle_0+\mathcal{C}
\ee
is the relaxation function and $\mathcal{C}$ a constant calculated via the final value theorem \cite{kubo2012}. Here, $\widetilde{\Psi}_0(t)\equiv \Psi_0(0)-\la\pd_{\lambda\lambda}^2\mc{H}\ra_0$. Combining Eqs. (\ref{eq:work}) and (\ref{eq:genforce-relax}), the average work performed at the linear response of the generalized force is
\begin{equation}
\begin{split}
W = &\, \delta\lambda\la\pd_{\lambda}\mc{H}\ra_0-\frac{\delta\lambda^2}{2}\widetilde{\Psi}_0\\
&+\delta\lambda^2 \int_0^\tau\int_0^t \Psi_0(t-t')\dot{g}(t')\dot{g}(t)dt'dt.
\label{eq:work2}
\end{split}
\end{equation}
We remark that in thermally isolated systems, the work is separated into two contributions: the quasistatic work $W_{\rm qs}$ and the excess work $W_{\rm ex}$. We observe that only the double integral on Eq.~(\ref{eq:work2}) has ``memory'' of the trajectory of $\lambda(t)$. Therefore the other terms are part of the contribution of the quasistatic work. Thus, we can split them as
\be
W_{\rm qs} = \delta\lambda\la\pd_{\lambda}\mc{H}\ra_0-\frac{\delta\lambda^2}{2}\widetilde{\Psi}_0,
\ee 
\begin{equation}
\begin{split}
W_{\text{ex}} = \frac{\delta\lambda^2}{2} \int_0^\tau\int_0^\tau \Psi_0(t-t')\dot{g}(t')\dot{g}(t)dt'dt.
\label{eq:wirrder}
\end{split}
\end{equation}
According the Ref.~\cite{naze2025fluctuation}, $W_{\rm ex}$ must be positive. Therefore, suitable systems to be described present the relaxation function with positive Fourier transform~\cite{naze2020compatibility}.

Finally, I remark that such treatment can be applied to classical systems, by changing the operators to functions, and the commutator by the Poisson bracket \cite{kubo2012}.

\section{Optimal excess work}

Consider the excess work rewritten in terms of the protocols $g(t)$ instead of its derivative
\textcolor{black}{\begin{equation}
\begin{split}
    W_{\rm ex} =& \frac{\delta\lambda^2}{2}\Psi_0(0)+\delta\lambda^2\int_0^\tau \dot{\Psi}_0(\tau-t)g(t)dt\\&-\frac{\delta\lambda^2}{2}\int_0^\tau\int_0^\tau \ddot{\Psi}_0(t-t')g(t)g(t')dt dt'.
    \end{split}
\end{equation}}
Observe that such a functional is identical to that optimized in Ref.~\cite{naze2024analytical}. To optimize in this context, I use the same procedure used previously: using calculus of variations, I derive the Euler-Lagrange equation that furnishes the optimal protocol $g^*(t)$ of the system that will minimize the excess work
\be
\int_0^\tau \ddot{\Psi}_0(t-t')g^*(t')dt' = \dot{\Psi}_0(\tau-t),
\label{eq:eleq}
\ee
\textcolor{black}{whose asterisk notation indicates here, and in any other quantity, its optimal value}. In particular, the optimal excess work will be \cite{naze2022optimal}
\be
W_{\rm ex}^* = \frac{\delta\lambda^2}{2}\Psi_0(0)+\frac{\delta\lambda^2}{2}\int_0^\tau \dot{\Psi}_0(\tau-t)g^*(t)dt.
\ee
The solution of Eq.~\eqref{eq:eleq} is~\cite{naze2024analytical}
\be
g^*(t) = \frac{a_{-2}t+a_{-1}}{a_{-2}\tau+2 a_{-1}}+\sum_{n=0}^{\infty}\frac{a_n (\delta^{(n)}(t)-\delta^{(n)}(\tau-t))}{a_{-2}\tau+2 a_{-1}},
\ee
with
\be
a_{n} = \left.\frac{d^n}{ds^n}\left[\frac{1}{s}\frac{1}{\mathcal{L}_s[\ddot{\Psi}_0(t)/\Psi_0(0)]}\right]\right|_{s=0},
\ee
for $n\ge 0$. In particular, the first terms are
\be
a_{-2}=1,\quad a_{-1}=\mathcal{L}_s\left\{\frac{\Psi(t)}{\Psi(0)}\right\}(0).
\ee
Here, $\mathcal{L}_s$ is the Laplace transform. Also, to preserve the time-reversal symmetry in the points $t=0$ and $t=\tau$, one must define $\delta^{(n)}(0)=\delta^{(n)}(\tau)=0$, for all $n\in\mathcal{N}$. The aim of this work is to fully characterize such an optimal protocol and its optimal work for thermally isolated system. To start, I characterize them properly.

\section{Thermally isolated system features}

To distinguish thermally isolated systems from open systems, I need to characterize them properly. A well-defined relaxation time $\tau_R$ is an important physical quantity that distinguishes them. Observing that the relaxation time for an open system is given by
\be
\tau_R = \mathcal{L}_s\left\{\frac{\Psi_0(t)}{\Psi_0(0)}\right\}(0),
\ee
the same concept could be used for thermally isolated systems. However, the only important term that matters is the constant term of the Fourier series of the relaxation function, which is an even function. Indeed, the other Laplace transform of the remaining terms are identically equal to zero, because they are cosines. Therefore, since the system does not equilibrate after the driving, a thermally isolated system is such that the constant part of its Fourier series of the relaxation function is zero. This implies $\tau_R=0$. 

But what does that mean exactly? First of all, calling this quantity ``relaxation time'' for thermally isolated systems is technically wrong, since the relaxation function does not decorrelate as systems performing the isothermal processes do with the heat bath after the driving ceases. This new timescale must have an interpretation that coincides with the relaxation time for isothermal processes and must be zero for thermally isolated systems. I propose the following interpretation: this new timescale, which I will call ``waiting time'', is the average time necessary for the system to achieve its final state when the driving has ceased. In this manner, systems performing isothermal processes will equilibrate with the heat bath, having, therefore, a positive waiting time, while the ones performing adiabatic processes are already in their final states once their processes are stopped, meaning that their waiting time is null. In particular, for isothermal processes, the waiting time will coincide with the relaxation time.

At this point, to unify such interpretation for both cases, the waiting time $\tau_w$ is defined
\be
\tau_w = \mathcal{L}_s\left\{\frac{\Psi_0(t)}{\Psi_0(0)}\right\}(0),
\ee
The analytical solution, valid for isothermal and adiabatic processes, becomes
\be
g^*(t) = \frac{t+\tau_w}{\tau+2\tau_w}+\sum_{n=0}^{\infty}\frac{a_n (\delta^{(n)}(t)-\delta^{(n)}(\tau-t))}{\tau+2 \tau_w}.
\ee
\textcolor{black}{As observed in Ref.~\cite{naze2024analytical}, the uniqueness of such a solution is not guaranteed. Multiparameter problem can be conceived as well, with an addition of a positive-definite Casimir-Onsager matrix~\cite{naze2025casimir} for relaxation functions and a system of Euler-Lagrange equations}. Finally, a sufficient condition for being a thermally isolated system is having its relaxation function periodic with period $T$, such that
\be
\int_0^T\Psi_0(t)dt=0.
\ee
In this manner, its relaxation time is also null. Let us see the consequences of this fact.

\section{Null excess work}

The first important consequence is the existence of a protocol whose excess work produced is null for all switching times. Remark that this is the best minimum possible~\cite{naze2025fluctuation}. Therefore,
\be
\int_0^\tau \dot{\Psi}_0(\tau-t)g^*(t)dt = -\Psi_0(0).
\label{eq:wexnull}
\ee
To prove this fact I am going to show that it is always possible to find coefficients $a_n$ such that the optimal protocol with $\tau_w=0$ reproduces Eq.~\eqref{eq:wexnull}. Indeed, applying such protocol on Eq.~\eqref{eq:wexnull}, one has
\be
\int_0^\tau\Psi_0(t)dt = \sum_{n=0}^\infty a_n\int_0^\tau \dot{\Psi}_0(t)(\delta^{(n)}(\tau-t)-\delta^{(n)}(t))dt,
\ee
or simplifying more
\be
\Theta_0(\tau)-\Theta_0(0) = -\sum_{n=0}^\infty (-1)^n a_n[\dot{\Psi}^{(n)}_0(\tau)+\dot{\Psi}^{(n)}_0(0)],
\ee
where $\Theta_0(t)$ is the integral of the relaxation function. Since $\Theta_0(t)$ is odd, $\Theta_0(0)=0$ and the left side has $a_{2n+1}=0$. Using $\dot{\Psi}^{(2n)}_0(0)=0$, we simplify more
\be
\Theta_0(\tau) = -\sum_{n=0}^\infty a_{2n}\dot{\Psi}^{(2n)}_0(\tau).
\ee
Expanding $\Theta_0(\tau)$ and $\dot{\Psi}^{(2n)}_0(\tau)$ in Fourier series, 
\be
\Theta_0(\tau) = \sum_{m=0}^{\infty}b_m\sin{\omega_m \tau},\quad \dot{\Psi}^{(2n)}_0(\tau) = -\sum_{m=0}^{\infty}c_{mn}\sin{\omega_m \tau},
\ee
one finally has
\be
b_m = \sum_{n=0}^{\infty}c_{mn}a_{2n},
\ee
which we suppose is a solvable linear system. Therefore, the optimal excess work is null for any switching time for thermally isolated systems. Let us prove now that these optimal protocols are shortcuts to adiabaticity.

\section{Conservation of the adiabatic invariant}

Shortcuts to adiabaticity are protocols to which the adiabatic invariant is conserved during the driving~\cite{acconcia2015degenerate,acconcia2015shortcuts}. To prove that this holds for our problem, let us define the excess power $\mc{P}_{\rm ex}(t)$ as
\be
W_{\rm ex}(\tau)=\int_0^\tau \mc{P}_{\rm ex}(t)dt.
\ee
To rule out the breakdown of the conservation of the adiabatic invariant, the excess power must be null at all instants of the driving. Indeed, this will avoid the appearance of non-equilibrium excitations on the system~\cite{acconcia2015shortcuts}. Since the optimal excess work is zero for all switching times, the optimal excess power is
\begin{align}
\mc{P}_{\rm ex}^*(t) &= \dot{W}_{\rm ex}^*(t)+\frac{W_{\rm ex}^*(0^+)}{\tau}\\
&= \dot{W}_{\rm ex}^*(t) \\
&= 0.    
\end{align}
Therefore, the analytical optimal protocols are shortcuts to adiabaticity for all switching times.

\section{Examples}

Let us see two typical examples of thermally isolated systems.

\subsection{Cosine relaxation function}

First I solve the optimal protocol problem of a typical relaxation function of thermally isolated systems performing an adiabatic process~\cite{acconcia2015degenerate,acconcia2015shortcuts,naze2023adiabatic}
\be
\Psi_0(t) = \Psi_0(0)\cos{\omega t}.
\ee
Observe first that the system is periodic, with the integral of its relaxation function null during the period. The terms $a_n$ will be
\be
a_{-2}=1,\quad a_{-1}=0,\quad a_{0}=1/\omega^2,
\ee
and
\be
a_{n}=0,
\ee
with $n\ge 1$. The optimal protocol will be
\be
g^*(t) = \frac{t}{\tau}+\frac{\delta(t)-\delta(\tau-t)}{\tau\omega^2}.
\ee
As predicted, the optimal protocol is a shortcut to adiabaticity for any switching time. Indeed, calculating the excess work, one has
\be
W_{\rm ex}(\tau)=0.
\ee

\subsection{Transverse-field quantum Ising chain}

The relaxation function of the transverse-field quantum Ising chain taken initially at equilibrium with $T=0$ and periodic boundary conditions is~\cite{naze2022kibble}
\be
\Psi_0(t)=\frac{16}{N}\sum_{n=1}^{N/2}\frac{J^2}{\epsilon^3(n)}\sin^2{\left(\left(\frac{2n-1}{N}\right)\pi\right)}\cos{\left(\frac{2\epsilon(n)}{\hbar}t\right)},
\ee
where
\be
\epsilon(n)=2\sqrt{J^2+\Gamma_0^2-2 J \Gamma_0 \cos{\left(\left(\frac{2n-1}{N}\right)\pi\right)}},
\ee
where $J$ is the coupling energy of the system, $\Gamma_0$ the initial magnetic field and $N$ is a even number of spins. The waiting time of such a system is given by
\be
\tau_w = \left\{\begin{array}{lr}
        \infty, & \text{for } N\rightarrow\infty,\, J=\Gamma_0\\
        0, & \text{otherwise},
        \end{array}\right.
\ee
that is, the system has a discontinuous phase transition in the critical point $J=\Gamma_0$ and at the thermodynamical limit ($N\rightarrow \infty$). When the system is in that particular case, I will say that it is under a quenching process.

\subsubsection{Non-quenching process}

\begin{figure}
    \centering
    \includegraphics[scale=0.5]{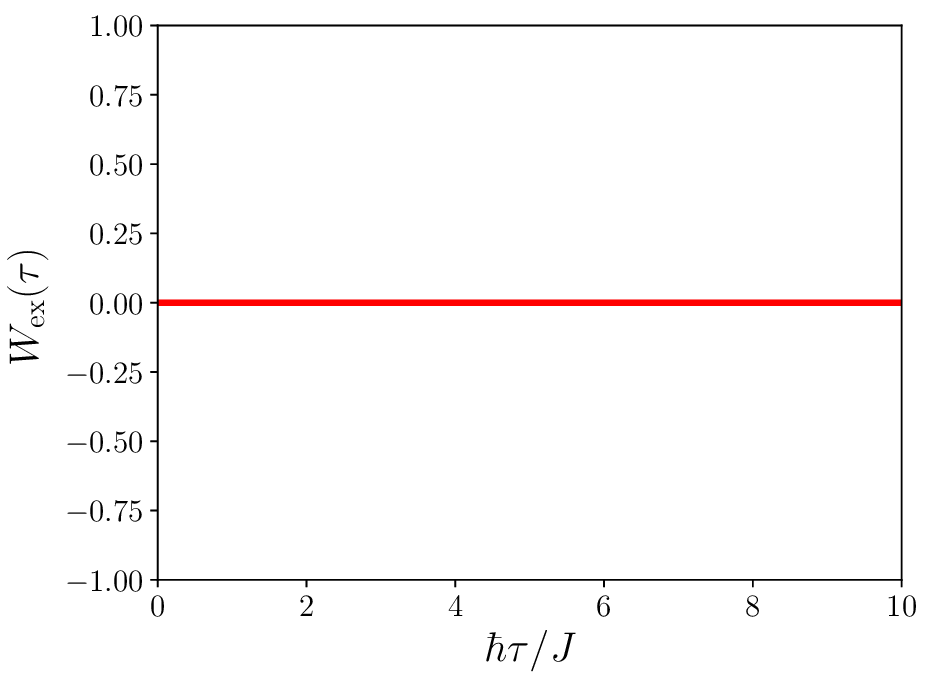}
    \caption{Optimal excess work calculated for the transverse-field quantum Ising chain under a non-quenching process. Indeed, it is a shortcut to adiabaticity for any switching time. It was used $\hbar=1,J=1,\Gamma_0=0.95,\delta\Gamma=0.1,N=10$.}
    \label{fig:wex}
\end{figure}

In that case, considering that the system is periodic with the integral of its relaxation function null, and the Laplace transform of the second derivative of the relaxation function of such a system is an odd function, the coefficients $a_n$ will be
\be
a_{-2}=1,\quad a_{-1}=0,\quad a_{2n}\neq 0, \quad a_{2n+1}=0,
\ee
with $n\ge 0$. Since $\tau_w=0$, the shortcut to adiabaticity will be
\be
g^*(t) = \frac{t}{\tau}+\sum_{n=0}^{\infty}\frac{a_{2n}(\delta^{(2n)}(t)-\delta^{(2n)}(\tau-t))}{\tau}.
\ee
To verify if it is indeed a shortcut, we calculate the excess work for driving with the following parameters: $\hbar=1,J=1, \Gamma_0=0.95, \delta\Gamma=0.1, N=10$. To avoid a calculation of an extensive series, we use the fact that the solution of the optimal protocol is global~\cite{naze2023global}, such that any solution found by alternative methods is an optimal protocol. \textcolor{black}{The procedure adopted is the following: choosing in the optimal protocol the number of Dirac delta and its derivatives equal to $2N$, I let the coefficients $a_{2n}$ free of choice. Therefore, the optimal protocol is
\be
g^*(t) = \frac{t}{\tau}+\frac{1}{\tau}\sum_{n=0}^{2N-1}a_{2n}(\delta^{(2n)}(t)-\delta^{(2n)}(\tau-t))
\label{eq:genoptprot}
\ee
I constructed a linear equation system of those coefficients from the Euler-Lagrange equation for each independent part of the sum (see Appendix~\ref{app:1} for details). Since the number of equations is $2N$, the system will have a solution. After solving it, for this example I found positive ``frequencies'' $\Omega_{2n}$ for each one of the derivatives of the Dirac delta, such that
\be
g^*(t) = \frac{t}{\tau}+\frac{1}{\tau}\sum_{n=0}^{2N-1}\frac{1}{\Omega_{2n}}(\delta^{(2n)}(t)-\delta^{(2n)}(\tau-t)).
\ee
The calculation of the excess work result for such a solution is depicted in Fig.~\ref{fig:wex}. Indeed, as predicted, we could suppress any excitation of the non-equilibrium driving completely. The situation changes dramatically with the quenching process case.}

\subsubsection{Quenching process}

In that particular case, the relaxation function of the system becomes proportional via a positive constant $c$ to $\Psi_0(0)$, which diverges
\be
\Psi_0(t) = c\Psi_0(0) \rightarrow \infty.
\ee
Because of this, for any finite-time protocol, the excess work will diverge
\be
W_{\rm ex}(\tau)=\frac{\delta\Gamma^2}{2}\Psi_0(0)\rightarrow \infty,
\ee
which is a manifestation of Kibble-Zurek mechanism~\cite{naze2022kibble,zurek1985cosmological, zurek1996cosmological, delcampo2014universality, polkovnikov2005universal, dziarmaga2005dynamics, dziarmaga2010dynamics, delcampo2010adiabatic, dziarmaga2006dynamics, cucchietti2007dynamics, zurek2009breaking}. Analyzing for the optimal case, I will indeed confirm the result. In that manner, for a finite-time process, the optimal protocol becomes
\be
g^*(t) = \frac{1}{2}+\sum_{n=0}^\infty b_n (\delta^{(n)}(t)-\delta^{(n)}(\tau-t)),
\ee
with
\be
b_n = \lim_{s\rightarrow 0^+}\frac{(-1)^{n}(n+1)!}{s^{n+1}}=(-1)^n\cdot\infty,
\ee
meaning that the optimality is achieved with a pausing protocol in the continuous part~\cite{marshall2019power, albash2020comparing, chen2022advantage, zielewski2022method, chen2023efficiency, chen2020why, calude2017adiabatic, marsh2019enhancing, passarelli2020reverse}. The optimal excess work becomes
\be
W_{\rm ex}(\tau) = \frac{\delta\Gamma^2}{4}(\Psi_0(\tau)+\Psi_0(0))\rightarrow\infty,
\ee
which diverges as well. In particular, the singular part does not offer any contribution, becoming the pausing protocol the only part important to obtain the diverging result. In this manner, optimizing the excess work in that particular regime is useless. Also, the optimal excess power is
\be
\mathcal{P}_{\rm ex}(t) = \frac{\delta\Gamma_0^2}{4}\dot{\Psi}_0(t)+\frac{\delta\Gamma_0^2}{2\tau}\Psi_0(0)=\frac{\delta\Gamma_0^2}{2\tau}\Psi_0(0)\rightarrow \infty,
\ee
indicating that the system has non-equilibrium excitations along the process.

Finally, remark that there is no paradox in the discontinuous phase transition presented in the waiting time and the continuous phase transition in Kibble-Zurek mechanism, as illustrated in Ref.~\cite{naze2022kibble} for linear response theory. The system passes indeed through a continuous phase transition, obtaining a diverging excess work for continuous approximation of $\Gamma_0$ to $J$ at the thermodynamical limit. However, at the same time, until the system does not achieve the critical point, it is always possible to suppress completely the non-equilibrium excitations of the system. The price to pay is to manage with singular functions as $N$ increases. Even though, this is excellent news about such a system, where the real-world scenario outperforms the ideal case in which Kibble-Zurek mechanism manifests itself. Such a result would surely benefit quantum annealing in adiabatic quantum computation.

\section{Discussion}

Let us see each case treated here concerning the singular and non-singular parts of the optimal protocol.

\subsection{Non-quenching process case}

In this case, $\tau_w<\infty$.

\subsubsection{Continuous part}

For all examples treated here, the continuous part of the shortcut to adiabaticity $g^*_C(t)$ was given by
\be
g^*_{C}(t)=\frac{t}{\tau},
\ee
which is independent of the waiting time, and always will be a straight line from $t=0$ to $t=\tau$. Also, this continuous part is restrained 
\be
0\le g^*_C(t)\le 1,
\ee
for all $t\in[0,\tau]$, $0\le\tau/\tau_w\le\infty$, meaning that it can be used in linear-response theory. Therefore, the shortcut is physically consistent.

\subsubsection{Singular part}

For all examples treated here, the singular part of the optimal protocol $g^*_S(t)$ was given by
\be
g^*_S(t)=\sum_{n=1}^{N/2}\frac{\delta^{(2(n-1))}(t)-\delta^{(2(n-1))}(\tau-t)}{\tau\Omega_n^{2n}},
\ee
where $\Omega_n$ were positive numbers, independent of $\tau$ and related to the parameters of the system. Therefore, for the extreme cases where $\tau\Omega_n^{2n}\gg 1$, one has
\be
\lim_{\tau\Omega_n^{2n}\gg 1}g^*_S(t)=0.
\ee
Also, for $\tau\Omega_n^{2n}\ll 1$, with $\tau\rightarrow 0^+$, the deltas and derivatives cancel out. Therefore
\be
\lim_{\tau\Omega_n^{2n}\ll 1}g^*_S(t)=0.
\ee
This result is coherent with the same one for open systems, where the sudden and slowly varying cases do not have a participative effect on the singular part of the solution. 

\subsection{Quenching process case}

In this case, $\tau_w=\infty$.

\subsubsection{Continuous part}

For finite-time process, where $\tau<\infty$, the continuous part of the optimal protocol $g^*_C(t)$ is given by
\be
g^*_{C}(t)=\frac{1}{2},
\ee
which is pausing protocol from $t=0$ to $t=\tau$. Also, this continuous part is restrained 
\be
0\le g^*_C(t)\le 1,
\ee
for all $t\in[0,\tau]$, $0\le\tau/\tau_w\le\infty$, meaning that it can be used in linear-response theory. Therefore, the optimal protocol is physically consistent in that sense.

\subsubsection{Singular part}

For finite-time processes, the singular part of the optimal protocol $g^*_S(t)$ is given by
\be
g^*_S(t)=\sum_{n=0}^\infty b_n(\delta^{(n)}(t)-\delta^{(n)}(\tau-t)),
\ee
with 
\be
b_n = (-1)^n\cdot\infty.
\ee
Observe that the singular part does not participate in the excess work, due to the relaxation function becoming a constant. In practice, only the pausing protocol deduced in the continuous part is essential to the protocol.

\subsection{Energetic cost of shortcut to adiabaticity}

\textcolor{black}{The analysis made above shows us that, in near-to-equilibrium processes, the optimal protocol reduces to the straight line. This indicates that the combination of Dirac deltas and their derivatives takes all the energetic cost along a finite-time driving. I propose that, as an energetic cost $\mathcal{C}$ of such shortcuts, the excess work of the singular part
\be
\mathcal{C} = \left|\frac{\delta\lambda^2}{2}\int_0^\tau\dot{\Psi}_0(\tau-t)g_S^*(t)dt\right|.
\ee
However, since the shortcut to adiabaticity nullifies the excess work, one can calculate such a quantity employing only the linear protocol
\be
\mathcal{C} = \left|-\frac{\delta\lambda^2}{2}\Psi_0(0)-\frac{\delta\lambda^2}{2}\int_0^\tau \dot{\Psi}_0(\tau-t)\frac{t}{\tau}dt\right|.
\ee
After some algebra, one has
\be
\mathcal{C} = \left|\frac{\delta\lambda^2}{2\tau}\int_0^\tau \Psi_0(t)dt\right|
\ee
Curiously, the energetic cost is proportional to the value of the time-averaged relaxation function calculated at the end of the process~\cite{naze2023adiabatic}. In the non-quenching case, for $\tau\rightarrow 0^+$, the energetic cost is 
\be
\mathcal{C} = \frac{\delta\lambda^2}{2}\Psi_0(0),
\ee
which is the value of the excess work in that limit. For $\tau\gg 1$, the energetic cost is zero, as expected. For the quenching case, the energetic cost is proportional to $\Psi_0(0)$, so it goes to infinity, showing the necessity of Dirac deltas and their derivatives to cope with these non-equilibrium dynamics.}

\section{Final remarks}
\label{sec:final} 

In this work, I presented an analytical solution for shortcuts to adiabaticity for any thermally isolated system performing a weakly driving process for any switching time. It relies only on the structure of the analytical solution of the optimal protocol of the irreversible work of open systems, \textcolor{black}{having a continuous part and a singular part, composed of Dirac deltas and their derivatives}. \textcolor{black}{Although it seems at odds the singular part of the solution, mainly from the experimental point-of-view, at least for Brownian particles modern optical waveform generators, such as those described in~\cite{yang2024compact}, reproduce with nice precision nascent delta as protocols}. Two examples are solved to illustrate the solution: The cosine relaxation function, which encompasses a variety of systems, and the transverse-field quantum Ising chain, a prototype of an adiabatic quantum computer. The concept of waiting time is introduced to unify the description of the analytical solution for isothermal and adiabatic processes. Also, although the result of the transverse-field quantum Ising chain suggests that errors that come from the excitability of non-equilibrium drivings can be completely suppressed when the critical point is not achieved, this is only valid when linear response theory holds, that is, for low $N$ and distant from the critical point~\cite{naze2022kibble}. \textcolor{black}{I verified as well how the solutions of the examples behave in extreme cases, like the slowing varying case and sudden case. In the end, an analysis of the energetic cost of these shortcuts of adiabaticity is made, where we consider the contribution of the singular part as the most important cost associated to the shortcut.}

\newpage

\appendix

\section{Optimal protocol of transverse-field quantum Ising chain}
\label{app:1}

For simplicity, let us call the relaxation function of the transverse-field quantum Ising chain as
\be
\Psi_0(t) = \sum_{n=1}^{N/2}A_n \cos{(\omega_n t)}.
\ee
Using the generic optimal protocol~\eqref{eq:genoptprot}, the Euler-Lagrange equation will be
\begin{equation}
\begin{split}
\sum_{n=1}^{N/2}&\sum_{m=0}^{2N-1}\int_0^\tau A_n a_m \omega_n^2 \cos{(\omega_n (t-t'))}[t'+(\delta^{(2m)}(t')\\&-\delta^{(2m)}(\tau-t'))]dt'=\sum_{n=1}^{N/2} \tau A_n \omega_n \sin{(\omega_n (\tau-t))}
\end{split}
\end{equation}
Both on the right and left-hand sides, there it will be $N$ terms of $\cos{(\omega_n t)}$ and $N$ terms of $\sin{(\omega_n t)}$. Equaling both sides and asking that the equation formed is equal to zero, there will be $2N$ equations. Also, the number of $a_n$ is $2N$, which guarantees the existence of the solution of the linear equation system.

\newpage

\bibliography{SALR}

\end{document}